\documentclass[]{aa}

\usepackage{txfonts}
\usepackage{graphicx} 
\usepackage{natbib} 
\usepackage{longtable} 
\usepackage{amssymb}
\usepackage{lscape}
\usepackage{rotating}
\usepackage[english]{babel}
\usepackage{grffile}
\usepackage{color}

\bibpunct{(}{)}{;}{a}{}{,}

\begin{document}

\title{The white dwarf population within 40~pc of the Sun}

\author{Santiago Torres\inst{1,2}\and
        Enrique Garc\'\i a--Berro\inst{1,2}}
        
\institute{Departament de F\'\i sica, 
           Universitat Polit\`ecnica de Catalunya, 
           c/Esteve Terrades 5, 
           08860 Castelldefels, 
           Spain
           \and
           Institute for Space Studies of Catalonia, 
           c/Gran Capita 2--4, 
           Edif. Nexus 201, 
           08034 Barcelona, 
           Spain}

\date{\today}

\titlerunning{The white dwarf population within 40~pc}
\authorrunning{S. Torres \& E. Garc\'\i a--Berro}

\offprints{E. Garc\'\i a--Berro}


\abstract 
          {The white dwarf luminosity function is an important tool to
            understand the properties of  the Solar neighborhood, like
            its star formation history, and its age.}
          {Here we present  a population synthesis study  of the white
            dwarf population  within 40~pc  from the Sun,  and compare
            the  results of  this  study with  the  properties of  the
            observed sample.}
          {We use  a state-of-the-art population synthesis  code based
            on  Monte Carlo  techniques,  that  incorporates the  most
            recent  and reliable  white  dwarf  cooling sequences,  an
            accurate description  of the Galactic neighborhood,  and a
            realistic treatment of all  the known observational biases
            and selection procedures.}
          {We find a good agreement between our theoretical models and
            the  observed   data.   In  particular,   our  simulations
            reproduce a  previously unexplained feature of  the bright
            branch of  the white  dwarf luminosity function,  which we
            argue is  due to a  recent episode of star  formation.  We
            also derive  the age  of the Solar  neighborhood employing
            the position  of the observed  cut-off of the  white dwarf
            luminosity function, obtaining $\sim 8.9\pm 0.2$~Gyr.}
          {We  conclude that  a detailed  description of  the ensemble
            properties of the population  of white dwarfs within 40~pc
            of the Sun allows us  to obtain interesting constraints on
            the history of the Solar neighborhood.}


\maketitle


\section{Introduction}

White  dwarfs  are  the  most common  stellar  evolutionary  end-point
\citep{review}.  Actually,  all stars  with masses smaller  than $\sim
10\,  M_{\sun}$ will  end  their lives  as  white dwarfs  \citep{GB97,
Poelarends2008}.  Hence, given the shape of the initial mass function,
the local population of white dwarfs carries crucial information about
the physical processes governing the evolution of the vast majority of
stars, and in particular of the total  amount of mass lost by low- and
intermediate-mas  stars  during the  red  giant  and asymptotic  giant
branch  evolutionary  phases. Also,  the  population  of white  dwarfs
carries  fundamental  information  about the  history,  structure  and
properties of the Solar neighborhood,  and specifically about its star
formation history  and age.   Clearly, obtaining all  this information
from the observed ensemble properties of the white dwarf population is
an important endeavour.

However,   to   obtain   useful    information   from   the   ensemble
characteristics of the white dwarf population three conditions must be
met.   Firstly, extensive  and  accurate observational  data sets  are
needed.   In particular,  individual spectra  of a  sufficiently large
number of white  dwarfs are needed.  This has  been possible recently,
with  the advent  of  large-scale automated  surveys, which  routinely
obtain reliable spectra for sizable samples of white dwarfs.  Examples
of these  surveys, although not the  only ones, are the  Sloan Digital
Sky   Survey   \citep{York},   and    the   SuperCOSMOS   Sky   Survey
\citep{Row2011}, which have allowed us to have extensive observational
data  for a  very large  number  of white  dwarfs. Secondly,  improved
models of  the atmospheres of white  dwarfs that allow to  model their
spectra  --  thus  granting  us unambiguous  determinations  of  their
atmospheric composition,  and accurate  measurements of  their surface
gravities and  effective temperatures  -- are  also needed.   Over the
last years, several  model atmosphere grids with  increasing levels of
detail  and  sophistication   have  been  released  \citep{Bergeron92,
Koester2001,  Kowalski06, Tremblay11,  Tremblay13}, thus  providing us
with  a consistent  framework  to analyze  the observational  results.
Finally, it  is also  essential to  have state-of-the-art  white dwarf
evolutionary sequences  to determine  their individual ages.   To this
regard, it is worth mentioning  that we now understand relatively well
the physics controlling the evolution of white dwarfs.  In particular,
it has  been known  for several  decades that  the evolution  of white
dwarfs  is  determined by  a  simple  gravothermal process.   However,
although  the basic  picture  of white  dwarf  evolution has  remained
unchanged  for some  time,  we  now have  very  reliable and  accurate
evolutionary tracks, which take into account all the relevant physical
process  involved in  their cooling,  and that  allow us  to determine
precise   ages   of    individual   white   dwarfs   \citep{Salaris10,
Renedo10}. Furthermore,  it is  worth emphasizing that  the individual
ages derived  in this way  are nowadays  as accurate as  main sequence
ages  \citep{Salaris}.   When all  these  conditions  are met,  useful
information  can be  obtained  from the  observed data.   Accordingly,
large  efforts have  been recently  invested in  successfully modeling
with a high degree of realism the observed properties of several white
dwarf populations,  like the Galactic  disk and  halo -- see  the very
recent works  of \cite{Cojocaru1} and \cite{Cojocaru2}  and references
therein --  and the  system of Galactic  open \citep{Garcia-Berro2010,
Bellini,   Bedin2010}   and  globular   clusters   \citep{Hansen_2002,
Gar_etal_2014, Torres2015}.

In this  paper we analyze the  properties of the local  sample of disk
white dwarfs, namely  the sample of stars with  distances smaller than
40~pc \citep{Limoges2013,  Limoges2015}. The most salient  features of
this sample  of white dwarfs are  discussed in Sect.~\ref{sec:obsdat}.
We  then  employ  a  Monte  Carlo  technique  to  model  the  observed
properties  of the  local sample  of  white dwarfs.   Our Monte  Carlo
simulator is  described in some detail  in Sect.~\ref{sec:popsyn}. The
results of  our populations  synthesis studies  are then  described in
Sect.~\ref{sec:res}.  In  this section we  discuss the effects  of the
selection criteria  (Sect.~\ref{subsec:selcri}), we calculate  the age
of the  Galactic disk (Sect.~\ref{subsec:fitage}), we  derive the star
formation      history       of      the       Solar      neighborhood
(Sect.~\ref{subsec:rburst}), and we determine  the sensitivity of this
age determination to the slope of the initial mass function and to the
adopted          initial-to-final           mass          relationship
(Sect.~\ref{subsec:effects}).   Finally,  in Sect.~\ref{sec:concl}  we
summarize our main results and we draw our conclusions.

\section{The observational sample}
\label{sec:obsdat}

Over the last decades several  surveys have provided us with different
samples of  disk white  dwarfs.  Hot  white dwarfs  are preferentially
detected using  ultraviolet color  excesses. The Palomar  Green Survey
\citep{Green86} and the Kiso  Schmidt Survey \citep{Kondo84} used this
technique to study  the population of hot disk  white dwarfs. However,
these surveys failed to probe the characteristics of the population of
faint, hence cool and redder, white dwarfs. Cool disk white dwarfs are
also normally  detected in  proper motion surveys  \citep{LDM88}, thus
allowing to  probe the faint  end of  the luminosity function,  and to
determine the  position of its  cut-off. Unfortunately, the  number of
white  dwarfs in  the faintest  luminosity bins  represents a  serious
problem.  Other recent magnitude-limited  surveys, like the SDSS, were
able to detect  many faint white dwarfs, thus allowing  to determine a
white  dwarf luminosity  function  which covers  the  entire range  of
interest   of   magnitudes,   namely    $7\la   M_{\rm   bol}\la   16$
\citep{Harris}.   However, the  sample  of  \cite{Harris} is  severely
affected  by the  observational biases,  completeness corrections  and
selection  procedures. \cite{Holberg08}  showed that  the best  way to
overcome these  observational drawbacks  is to rely  on volume-limited
samples.    Accordingly,  \cite{Holberg08}   and  \cite{Gianmichele12}
studied  the white  dwarf  population  within 20~pc  of  the Sun,  and
measured  the properties  of an  unbiased sample  of $\sim  130$ white
dwarfs.   The completeness  of  their samples  is  $\sim 90\%$.   More
recently, \cite{Limoges2013}  and \cite{Limoges2015} have  derived the
ensemble  properties of  a sample  of $\sim  500$ white  dwarfs within
40~pc of the Sun, using the  results of the SUPERBLINK survey, that is
a survey  of stars  with proper  motions larger  than 40~mas~yr$^{-1}$
\citep{Lepine2005}.   The estimated  completeness of  the white  dwarf
sample derived  from this survey is  $\sim 70\%$, thus allowing  for a
meaningful statistical  analysis. We will  compare the results  of our
theoretical  simulations  with  the white  dwarf  luminosity  function
derived  from this  sample.   However, a  few  cautionary remarks  are
necessary.  First, this  luminosity function has been  obtained from a
spectroscopic  survey that  has not  been yet  completed. Second,  the
photometry  is  not  yet  optimal.   Finally,  and  most  importantly,
trigonometric parallaxes are not available  for most cool white dwarfs
in  the  sample,  preventing accurate  determinations  of  atmospheric
parameters and radii  (hence, masses) of each  individual white dwarf.
For these  stars \cite{Limoges2015}  were forced to  assume a  mass of
$0.6\,   M_{\sun}$.   All   in   all,  the   luminosity  function   of
\cite{Limoges2015} is still somewhat  preliminary, but nevertheless is
the  only  one based  on  a  volume-limited  sample extending  out  to
40~pc. Nevertheless, we explore the  effects of these issues below, in
Sects.~\ref{subsec:selcri} and \ref{subsec:fitage}.

\section{The population synthesis code}
\label{sec:popsyn}

A detailed description  of the main ingredients employed  in our Monte
Carlo population  synthesis code  can be found  in our  previous works
\citep{Gar1999,Tor2001,   Tor2002,Gar2004}.    Nevertheless,  in   the
interest  of  completeness,  here  we  summarize  its  main  important
features.   

The  simulations described  below  were done  using  the generator  of
random  numbers  of  \cite{James_1990}.   This  algorithm  provides  a
uniform probability densities within  the interval $(0,1)$, ensuring a
repetition period  of $\ga 10^{18}$, which  for practical applications
is  virtually infinite.   For  Gaussian  probability distributions  we
employed  the  Box-Muller  algorithm  \citep{NRs}.  For  each  of  the
synthetic white  dwarf populations described below,  we generated $50$
independent Monte Carlo simulations employing different initial seeds.
Furthermore, for each of these  Monte Carlo realizations, we increased
the  number of  simulated  Monte Carlo  realizations  to $10^4$  using
bootstrap techniques --  see \cite{Cam2014} for details.   In this way
convergence in all the final values of the relevant quantities, can be
ensured. In the  next sections we present the ensemble  average of the
different Monte Carlo  realizations for each quantity  of interest, as
well as the corresponding standard deviation. Finally, we mention that
the  total  number  of  synthetic  stars  of  the  restricted  samples
described below and  the observed sample are always  similar.  In this
way  we  guarantee that  the  comparison  of  both  sets of  data  are
statistically sound.

To produce  a consistent white  dwarf population we first  generated a
set  of random  positions of  synthetic  white dwarfs  in a  spherical
region centered on  the Sun, adopting a radius of  $50$~pc.  We used a
double exponential  distribution for the  local density of  stars. For
this density distribution we adopted  a constant Galactic scale height
of 250~pc  and a constant  scale length  of 3.5~kpc.  For  our initial
model the  time at which  each synthetic  star was born  was generated
according to a constant star formation rate, and adopting a age of the
Galactic disk  age, $t_{\rm disk}$.  The  mass of each star  was drawn
according to  a Salpeter mass function  \citep{Salpeter} with exponent
$\alpha=2.35$, except  otherwise stated,  which is  totally equivalent
for the relevant range of masses to the standard initial mass function
of \cite{Kroupa_2001}.   Velocities were  randomly chosen  taking into
account the differential rotation of the Galaxy, the peculiar velocity
of  the Sun  and a  dispersion law  that depends  on the  scale height
\citep{Gar1999}.  The evolutionary ages  of the progenitors were those
of  \cite{Renedo10}.   Given  the  age  of the  Galaxy  and  the  age,
metallicity, and mass of the  progenitor star, we know which synthetic
stars have had  time to become white dwarfs, and  for these, we derive
their   mass   using   the    initial-final   mass   relationship   of
\cite{Cat2008},  except otherwise  stated. We  also assign  a spectral
type  to each  of  the artificial  stars. In  particular,  we adopt  a
fraction of 20\% of  white dwarfs with hydrogen-deficient atmospheres,
while the rest of stars is assumed to be of the DA spectral type.

The set  of adopted  cooling sequences  employed here  encompasses the
most  recent  evolutionary  calculations  for  different  white  dwarf
masses.   For white  dwarf  masses smaller  than  $1.1\, M_{\sun}$  we
adopted the cooling tracks of white dwarfs with carbon-oxygen cores of
\cite{Renedo10}  for stars  with  hydrogen  dominated atmospheres  and
those of \cite{DBs} for hydrogen-deficient envelopes.  For white dwarf
masses larger  than this  value we used  the evolutionary  results for
oxygen-neon  white   dwarfs  of  \cite{Alt2007}   and  \cite{Alt2005}.
Finally, we  interpolated the  luminosity, effective  temperature, and
the value of  $\log g$ of each synthetic star  using the corresponding
white dwarf evolutionary tracks  .  Additionally, we also interpolated
their $UBVRI$  colors, which  we then converted  to the  $ugriz$ color
system.

\section{Results}
\label{sec:res}

\subsection{The effects of the selection criteria}
\label{subsec:selcri}

\begin{figure}[t]
\begin{center}
   {\includegraphics[trim = 10mm 35mm 10mm 35mm, clip, width=\columnwidth]{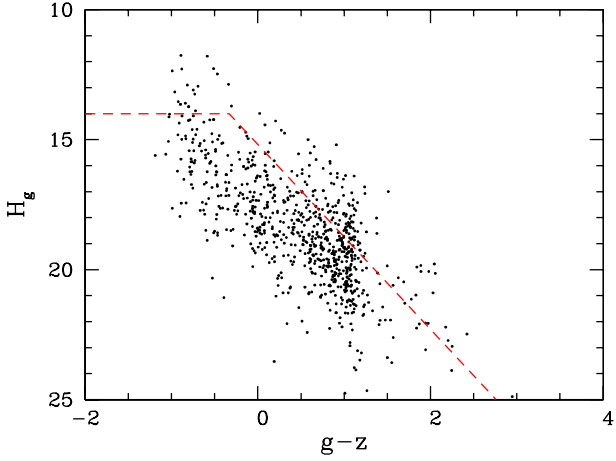}}
   {\includegraphics[trim = 10mm 35mm 10mm 35mm, clip, width=\columnwidth]{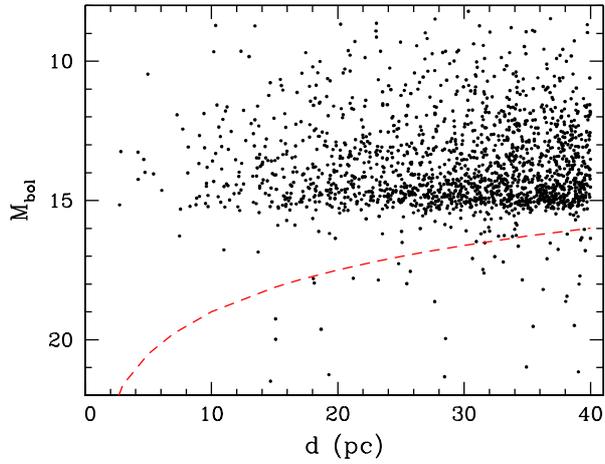}}
\end{center}
   \caption{Top panel:  Effects of  the reduced proper  motion diagram
     cut on  the synthetic population  of white dwarfs.  Bottom panel:
     Effects of the cut in $V$  magnitude in the simulated white dwarf
     population. In both  panels the synthetic white  dwarfs are shown
     as  solid  dots,  whereas  the red  dashed  lines  represent  the
     selection cut.}
\label{f:sel_two}
\end{figure}

A  key point  in the  comparison of  a synthetic  population of  white
dwarfs  with   the  observed  data   is  the  implementation   of  the
observational  selection  criteria  in the  theoretical  samples.   To
account for the observational biases with a high degree of fidelity we
implemented  in  a  strict  way the  selection  criteria  employed  by
\cite{Limoges2013,Limoges2015}  in their  analysis  of the  SUPERBLINK
database.  Specifically,  we only  considered objects in  the northern
hemisphere ($\delta>0^{\circ}$)  up to a  distance of 40~pc,  and with
proper  motions larger  than  $\mu>40\,{\rm  mas\,yr^{-1}}$. Then,  we
introduced a cut in the reduced  proper motion diagram $(H_g, g-z)$ as
\cite{Limoges2013} did  -- see  their Fig.~1  -- eliminating  from the
synthetic    sample   of    white    dwarfs    those   objects    with
$H_g>3.56(g-z)+15.2$  that are  outside of  location where  presumably
white  dwarfs   should  be   found.   Finally,   we  only   took  into
consideration  those  stars  with  magnitudes  brighter  than  $V=19$.

\begin{figure}[t]
\begin{center}
   {\includegraphics[trim = 10mm 35mm 10mm 35mm, clip, width=\columnwidth]{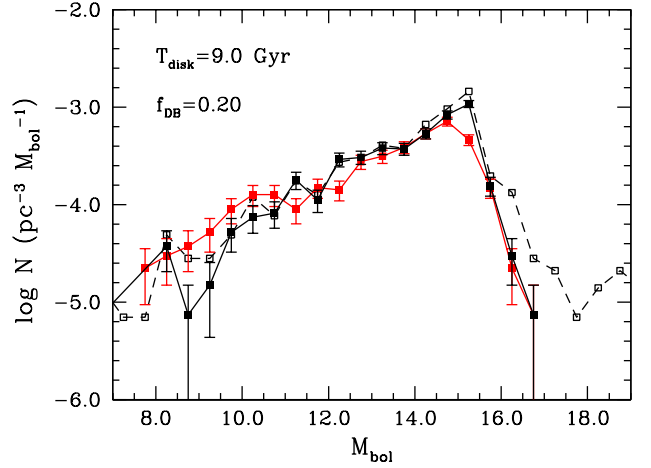}}
   {\includegraphics[trim = 10mm 35mm 10mm 35mm, clip, width=\columnwidth]{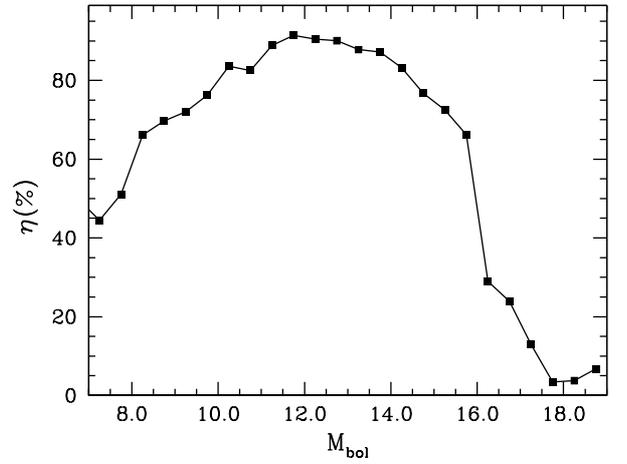}}
\end{center}     
   \caption{Top  panel:  Synthetic  white dwarf  luminosity  functions
     (black lines)  compared to the observed  luminosity function (red
     line).   The solid  line  shows the  luminosity  function of  the
     simulated white dwarf population  when all the selection criteria
     have  been  considered,  while   the  dashed  line  displays  the
     luminosity  function   of  the  entire  sample.    Bottom  panel:
     completeness of white dwarf population for our reference model.}
\label{f:triple}
\end{figure}

In Fig.~\ref{f:sel_two} we show the effects  of these last two cuts on
the  entire population  of white  dwarfs for  our fiducial  model.  In
particular, in the top panel of  this figure the reduced proper motion
diagram $(H_g, g-z)$ of the  theoretical white dwarf population (black
dots) and the  corresponding selection criteria (red  dashed line) are
displayed.   As can  be seen,  the  overall effect  of this  selection
criterion is that the selected sample  is, on average, redder than the
population from which it is drawn, independently of the adopted age of
the disk.   Additionally, in the bottom  panel of Fig.~\ref{f:sel_two}
we plot the bolometric magnitudes of  the individual white dwarfs as a
function of their  distance for the synthetic  white dwarf population.
The  red  dashed  line  represents the  selection  cut  in  magnitude,
$V=19$.  It  is clear  that  this  cut  eliminates faint  and  distant
objects. Also, it is evident that the number of synthetic white dwarfs
increases   smoothly  for   increasing   magnitudes   up  to   $M_{\rm
bol}\approx15.0$, and that for magnitudes larger than this value there
is a dramatic drop in the white dwarf number counts.  Furthermore, for
distances  of  $\sim  40$~pc  the  observational  magnitude  cut  will
eliminate  all white  dwarfs  with bolometric  magnitudes larger  than
$M_{\rm bol}\approx16.0$.  However, this magnitude cut still allows to
resolve the  sharp drop-off in  the number  counts of white  dwarfs at
magnitude $M_{\rm bol}\approx15.0$.  This, in turn, is important since
as it will be shown below  will allow us to unambigously determine the
age of the Galactic disk.

\begin{figure}[t]
\begin{center}
   {\includegraphics[trim = 10mm 35mm 10mm 35mm, clip, width=\columnwidth]{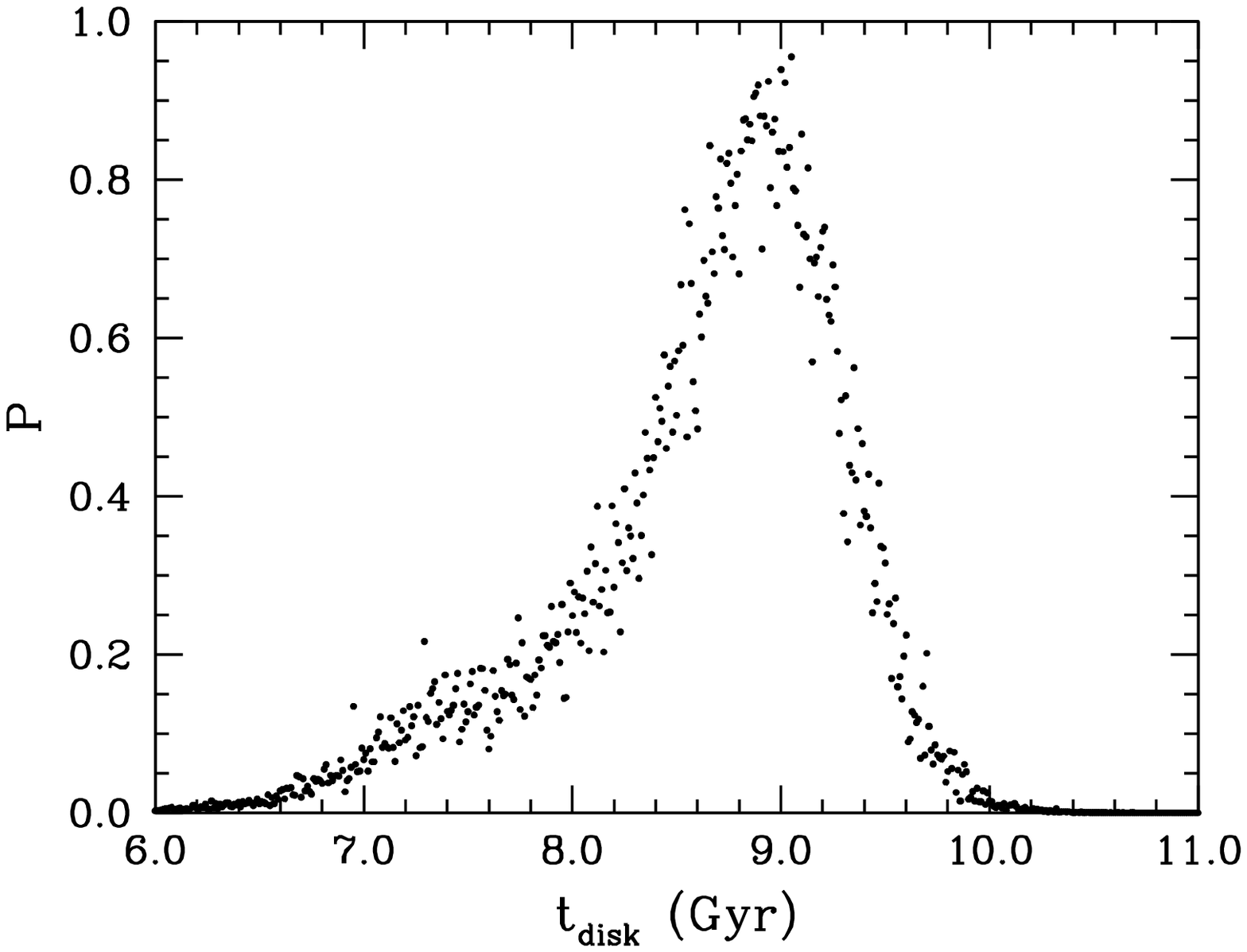}}
   {\includegraphics[trim = 10mm 35mm 10mm 35mm, clip, width=\columnwidth]{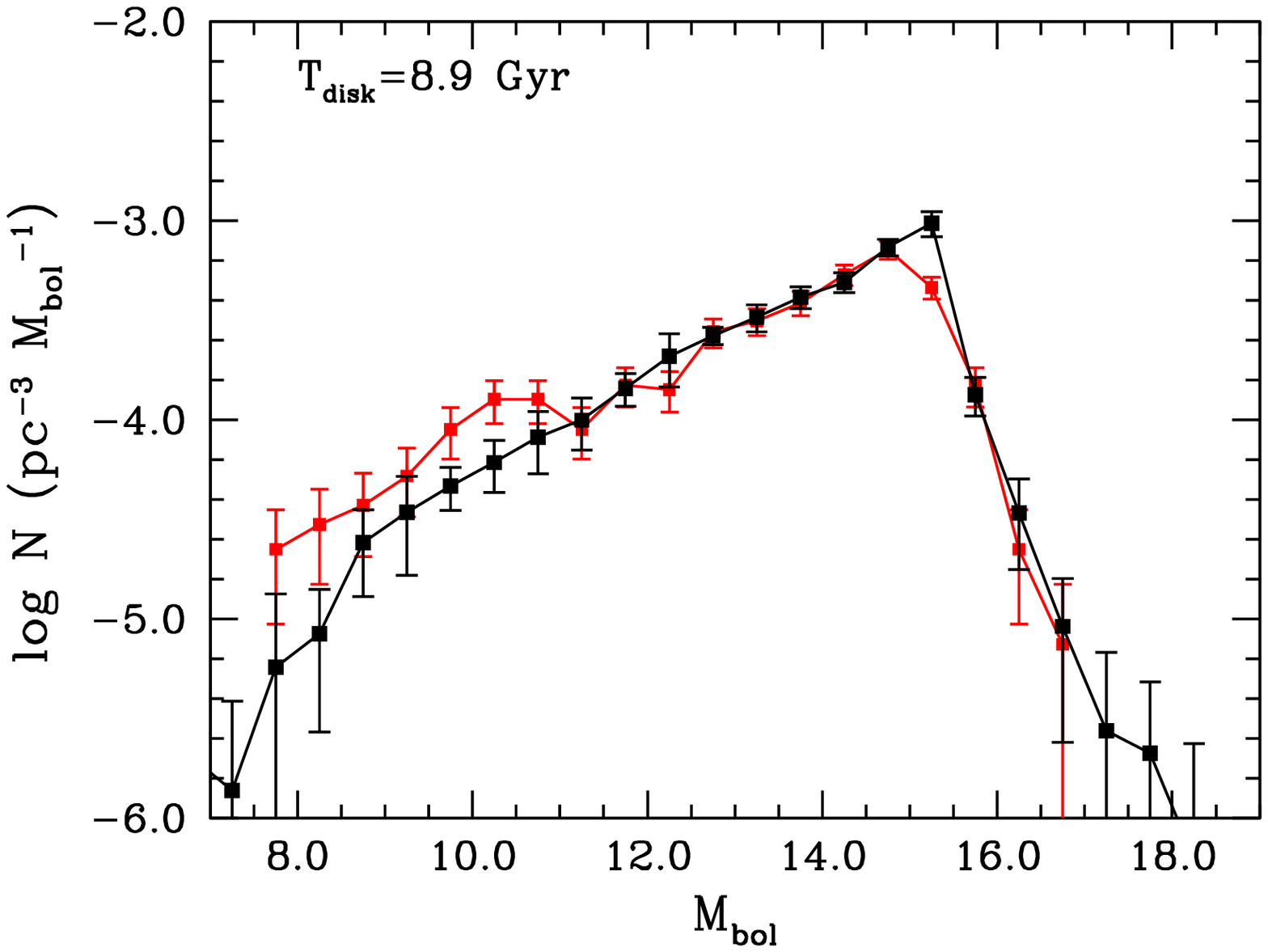}}
\end{center}
   \caption{Top panel: $\chi^2$ probability test  as a function of the
     age  obtained by  fitting the  three faintest  bins defining  the
     cut-off  of the  white dwarf  luminosity function.  Bottom panel:
     white dwarf luminosity function for the best-fit age.}
\label{f:chi_age}
\end{figure}

\begin{figure}[t]
\begin{center}
   {\includegraphics[trim = 10mm 35mm 10mm 35mm, clip, width=\columnwidth]{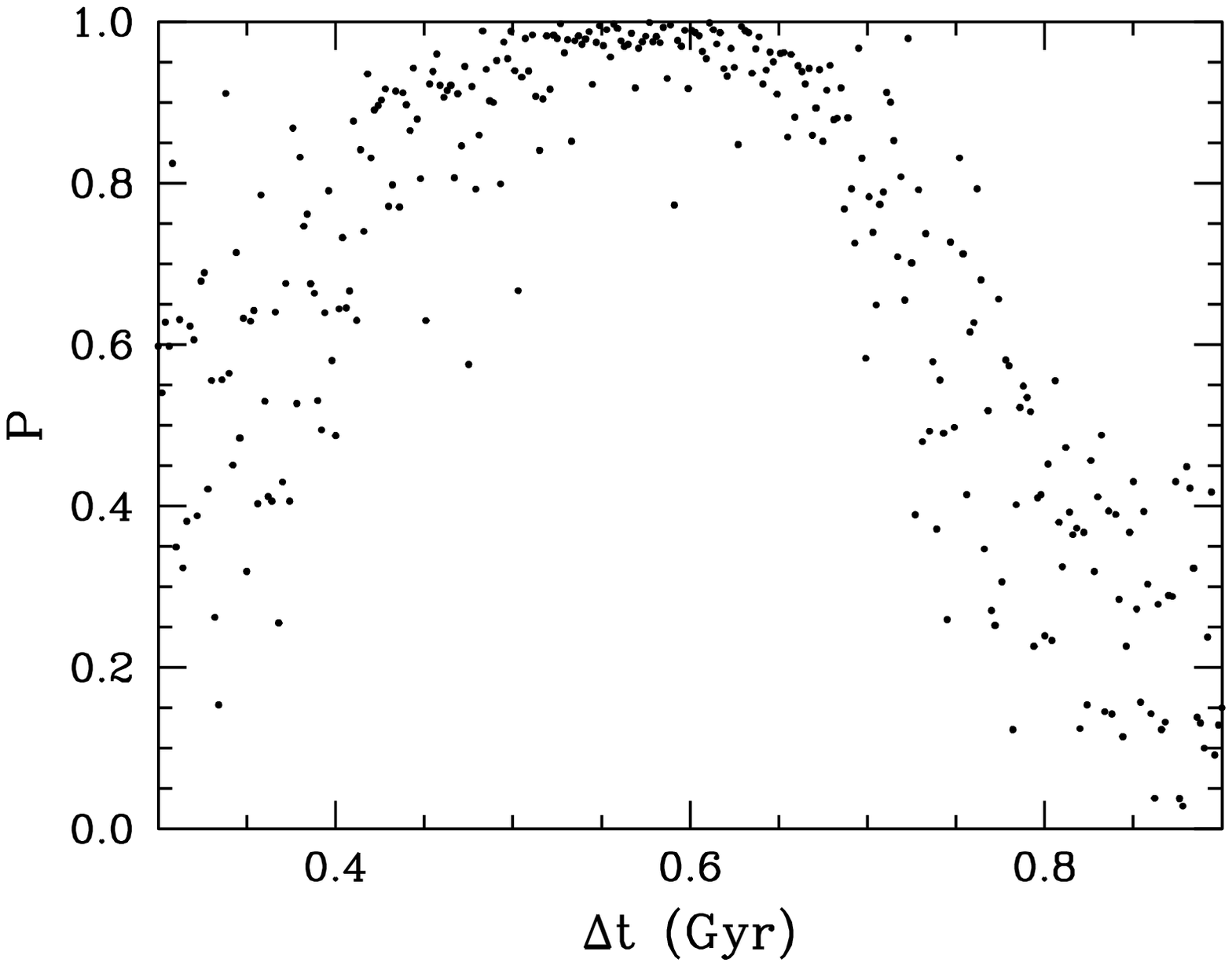}}
   {\includegraphics[trim = 10mm 35mm 10mm 35mm, clip, width=\columnwidth]{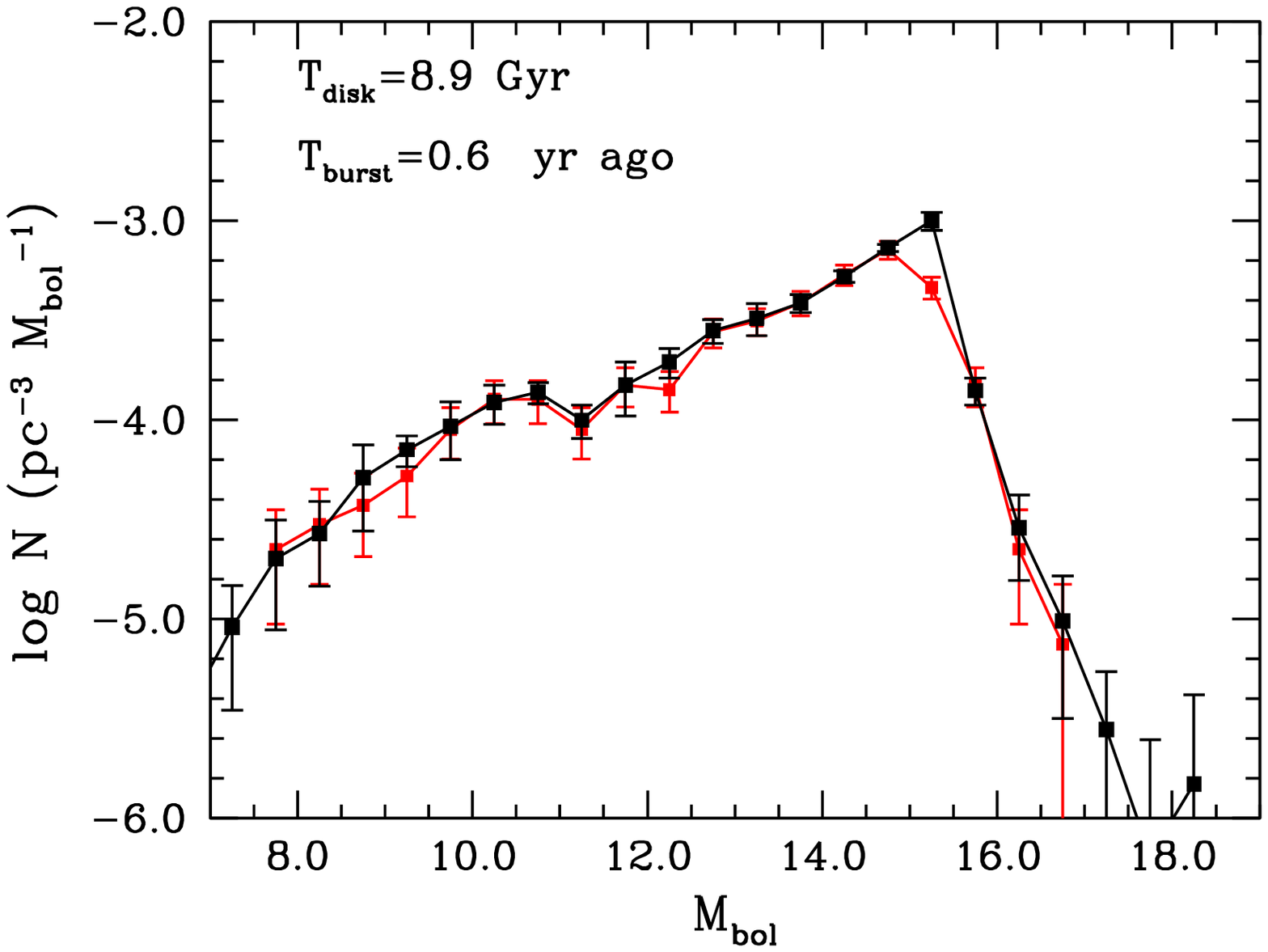}}
\end{center}
   \caption{Top panel: $\chi^2$ probability test  as a function of the
     age of the  burst of star formation obtained by  fitting the nine
     brightest  bins of  the white  dwarf luminosity  function. Bottom
     panel: Synthetic white  dwarf luminosity function for  a disk age
     of $8.9\,$Gyr and a recent burst of start formation (black line),
     compared  with the  observed white  dwarf luminosity  function of
     (red lines).}
\label{f:wdlf_burst}
\end{figure}

We  now study  if our  modeling of  the selection  criteria is  robust
enough. This is an important issue because reliable $ugriz$ photometry
was    available   only    for    a   subset    of   the    SUPERBLINK
catalog. Consequently,  \cite{Limoges2015} used photometric  data from
other   sources   like   2MASS,    Galex,   and   USNO-B1.0   --   see
\cite{Limoges2013} for details.  It is unclear how  this procedure may
affect the  observed sample.   Obviously, simulating all  the specific
observational procedures  is too  complicated for  the purpose  of the
present  analysis,   but  we  conducted  two   supplementary  sets  of
simulations to assess the reliability of our results.  In the first of
these sets we discarded an additional  fraction of white dwarfs in the
theoretical samples obtained after applying all the selection criteria
previously  described.  We  found that  if the  fraction of  discarded
synthetic  stars is  $\la  15\%$ the  results  described below  remain
unaffected.  Additionally, in a second  set of simulations we explored
the possibility that the sample of \cite{Limoges2015} is indeed larger
than  that  used  to  compute  the  theoretical  luminosity  function.
Accordingly, we  artificially increased the number  of synthetic white
dwarfs which  pass the  successive selection  criteria in  the reduced
proper motion diagram.  In particular we increased by  15\% the number
of artificial  white dwarfs populating  the lowest luminosity  bins of
the  luminosity function  (those  with $M_{\rm  bol}>12$).  Again,  we
found that  the differences  between both sets  of simulations  -- our
reference simulation and this one -- are minor.

\begin{figure*}[t]
\begin{center}
   {\includegraphics[trim = 0mm 34mm 0mm 10mm, clip, width=0.8\textwidth]{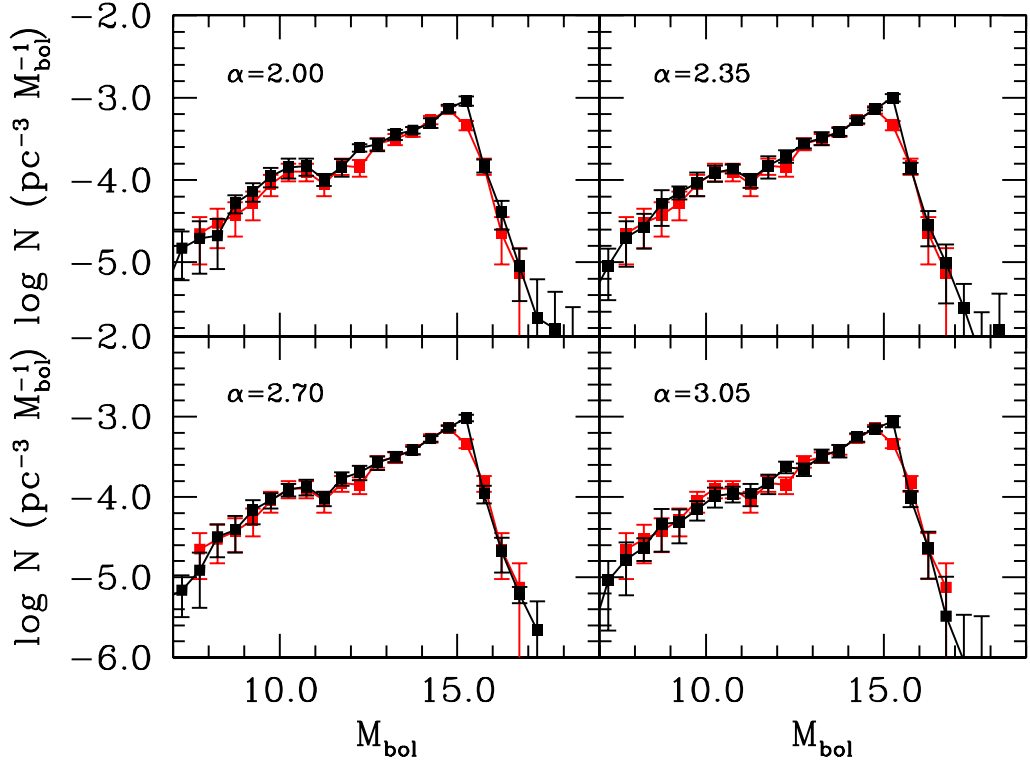}}
\end{center}
   \caption{Synthetic white  dwarf luminosity function for  a disk age
     of $8.9$~Gyr and a recent  burst of start formation for different
     values of the  slope of the Salpeter IMF  (black lines), compared
     with   the   observed   white  dwarf   luminosity   function   of
     \cite{Limoges2015} -- red line.}
\label{f:wdlf_alfa}
\end{figure*}

Once the  effects of the  observational biases and  selection criteria
have been analyzed, a theoretical  white dwarf luminosity function can
be built and  compared to the observed one. To  allow for a meaningful
comparison between  the theoretical and the  observational results, we
grouped  the synthetic  white  dwarfs using  the  same magnitude  bins
employed  by  \cite{Limoges2015}.   We emphasize  that  the  procedure
employed by  \cite{Limoges2015} to  derive the white  dwarf luminosity
function  simply consists  in counting  the  number of  stars in  each
magnitude bin, given that their sample is volume limited.  That is, in
principle, their number counts should  correspond with the true number
density  of  objects  per  bolometric magnitude  and  unit  volume  --
provided that their sample is complete without the need for correcting
the number counts using the $1/V_{\rm max}$ method -- or an equivalent
method  --  as it  occurs  for  magnitude  and proper  motion  limited
samples.

In the  top panel of  Fig.~\ref{f:triple} (top panel)  the theoretical
results are  shown using  black lines,  while the  observed luminosity
function  is  displayed  using  a red  line.   Specifically,  for  our
reference model we show the number of white dwarfs per unit bolometric
magnitude and volume for the entire theoretical white dwarf population
when  no selection  criteria  are  employed --  dashed  line and  open
squares --  and the  luminosity function  obtained when  the selection
criteria  previously  described are  used  --  solid line  and  filled
squares.   It  is worthwhile  to  mention  here that  the  theoretical
luminosity functions  have been  normalized to  the bin  of bolometric
magnitude $M_{\rm  bol}=14.75$ which corresponds to  the magnitude bin
for  which  the  observed  white dwarf  luminosity  function  has  the
smallest error bars. Thus, since this  luminosity bin is very close to
the maximum of the luminosity function, the normalization criterion is
practically equivalent  to a number density  normalization. As clearly
seen  in this  figure  the  theoretical results  match  very well  the
observed data, except for a quite apparent excess of hot white dwarfs,
which  will be  discussed  in detail  below.  Note  as  well that  the
selection criteria employed by \cite{Limoges2015} basically affect the
low luminosity  tail of the  white dwarf luminosity function,  but not
the location of the observed drop-off in the white dwarf number counts
nor that of the maximum of  the luminosity function.  This can be more
easily seen  by looking  at the  bottom panel  of Fig.~\ref{f:triple},
where the  completeness of the  simulated restricted sample  is shown.
We  found  that the  completeness  of  the  entire sample  is  $78\%$.
However,  the restricted  sample  is nearly  complete at  intermediate
bolometric magnitudes -- between $M_{\rm  bol}\simeq 10$ and 15 -- but
decreases very rapidly for magnitudes  larger than $M_{\rm bol}=15$, a
clear    effect   of    the    selection    procedure   employed    by
\cite{Limoges2015}.    Nevertheless,  this   low-luminosity  tail   is
populated  preferentially  by  helium-atmosphere stars,  and  by  very
massive oxygen-neon white dwarfs.  The prevalence of helium-atmosphere
white dwarfs  at low luminosities is  due to the fact  that stars with
hydrogen-deficient  atmospheres  have  lower luminosities  than  their
hydrogen-rich counterparts of the same  mass and age, because in their
atmospheres collision-induced  absorption does not play  a significant
role and cool to a very good approximation as black bodies.  Also, the
presence of massive oxygen-neon white dwarfs is a consequence of their
enhanced cooling rate, due to their smaller heat capacity.

\subsection{Fitting the age}
\label{subsec:fitage}

Now  we estimate  the age  of the  disk using  the standard  method of
fitting  the position  of the  cut-off of  the white  dwarf luminosity
function.  We  did this  by comparing  the faint  end of  the observed
white  dwarf   luminosity  function  with  our   synthetic  luminosity
functions.  Despite  the fact  that the  completeness of  the faintest
bins of the luminosity function  is substantially smaller (below $\sim
60\%$), we demonstrated  in the previous section that  the position of
the   cut-off    remains   nearly   unaffected   by    the   selection
procedures. Accordingly, we ran a set of Monte Carlo simulations for a
wide range of disk ages. We then  employed a $\chi^2$ test in which we
compared the theoretical and observed number counts of those bins that
define the  cut-off, namely  the three last  bins (those  with $M_{\rm
bol}>15.5$)  of  the  luminosity  function.    In  the  top  panel  of
Fig.~\ref{f:chi_age} we  plot this  probability as  a function  of the
disk age.  The best  fit is obtained for an age  of $8.9$~Gyr, and the
width of  the distribution at  half-maximum is $0.4$~Gyr.   The bottom
panel of this figure shows the white dwarf luminosity function for the
best-fit age.

One possible concern  could be that the age derived  in this way could
be affected by  the assumption that the mass of  cool white dwarfs for
which  no trigonometric  parallax  could be  measured was  arbitrarily
assumed to  be $0.6\, M_{\sun}$.  This may have  an impact of  the age
determination  using  the  cut-off   of  the  white  dwarf  luminosity
function. To assess  this issue we conducted  an additional simulation
in which  all synthetic  white dwarfs with  cooling times  longer than
1~Gyr have  this mass.  We  then computed the new  luminosity function
and derived the corresponding age  estimate.  We found that difference
of ages between both calculations is smaller than 0.1~Gyr.

\subsection{A recent burst of star formation}
\label{subsec:rburst}

As  clearly shown  in  the bottom  panel  of Fig.~\ref{f:chi_age}  the
agreement between the theoretical simulations and the observed results
is  very  good except  for  the  brightest  bins  of the  white  dwarf
luminosity,  namely   those  with  $M_{\rm  bol}\la   11$.  Also,  our
simulations fail to reproduce the shape  of the peak of the luminosity
function,  an   aspect  which  we   investigate  in  more   detail  in
Sect.~\ref{subsec:effects}.   The  excess  of  white  dwarfs  for  the
brightest  luminosity bins  is statistically  significant, as  already
noted  by  \cite{Limoges2015}.  \cite{Limoges2015}  already  discussed
various possibilities and pointed out that the most likely one is that
this feature of the white dwarf  luminosity function might be due to a
recent burst of star formation.  \cite{NohScalo1990} demonstrated some
time ago that  a burst of star formation generally  produces a bump in
the luminosity function, and that the  position of the bump on the hot
branch of the luminosity function is ultimately dictated by the age of
the burst of star formation -- see also \cite{Rowell13}.

\begin{figure}[t]
\begin{center}
   {\includegraphics[trim = 10mm 35mm 12mm 30mm, clip, width=0.95\columnwidth]{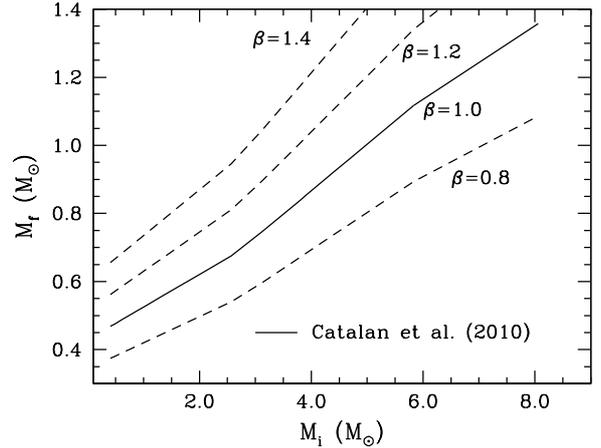}}
\end{center}
\caption{Initial-final mass  relationships adopted  in this  work. The
  solid line shows the  semi-empirical initial-final mass relationship
  of \cite{Catalan2008}, while the dashed  lines have been obtained by
  multiplying the final white dwarf mass by a constant factor $\beta$,
  as labeled.}
\label{f:mimf}
\end{figure}

\begin{figure*}[t]
\begin{center}
   {\includegraphics[trim = 0mm 34mm 0mm 10mm, clip, width=0.8\textwidth]{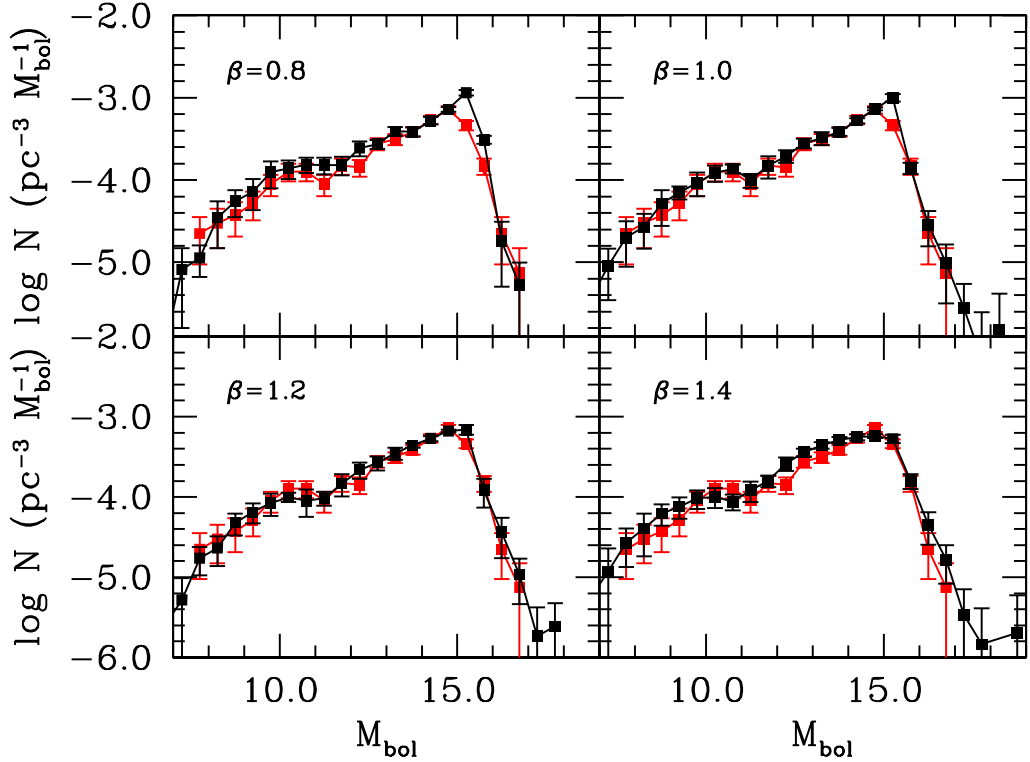}}
\end{center}
   \caption{Synthetic white  dwarf luminosity function for  a disk age
     of $8.9\,$Gyr and  a recent burst of start  formation for several
     choices of  the initial-to-final mass relationship  (black lines)
     compared  with the  observed white  dwarf luminosity  function of
     \cite{Limoges2015} -- red lines. See text for details.}
\label{f:wdlf_beta}
\end{figure*}

According to  these considerations, we  explored the possibility  of a
recent burst of star formation by  adopting a burst that occurred some
time ago and stays active until present.  The strength of this episode
of star  formation is another parameter  that can be varied.   We thus
ran  our Monte  Carlo  simulator using  a  fixed age  of  the disk  of
$8.9\,$Gyr and considered the time  elapsed since the beginning of the
burst, $\Delta t$, and its strength as adjustable parameters.  The top
panel  of Fig.~\ref{f:wdlf_burst}  shows the  probability distribution
for $\Delta t$,  computed using the same procedure  employed to derive
the age  of the  Solar neighborhood, but  adopting the  nine brightest
bins of the  white dwarf luminosity function, which  correspond to the
location of the bump of the white dwarf luminosity function.  The best
fit is  obtained for a burst  that happened $\sim 0.6\pm  0.2$~Gyr ago
and is $\sim  5$ times stronger that the constant  star formation rate
adopted in  the previous section.  As  can be seen in  this figure the
probability  distribution  function does  not  have  a clear  gaussian
shape. Moreover, the maximum of  the probability distribution is flat,
and  the  dispersion   is  rather  high,  meaning   that  the  current
observational data set does not allow to constrain in an effective way
the properties of this episode  of star formation.  However, when this
episode  of  star  formation  is  included  in  the  calculations  the
agreement between  the theoretical calculations and  the observational
results is  excellent. This is  clearly shown  in the bottom  panel of
Fig.~\ref{f:wdlf_burst}, where we show our best fit model, and compare
it   with   the   observed   white  dwarf   luminosity   function   of
\cite{Limoges2015}. As can  be seen, the observed excess  of hot white
dwarfs is now perfectly reproduced by the theoretical calculations.

\subsection{Sensitivity of the age to the inputs}
\label{subsec:effects}

In this section we will study the sensitivity of the age determination
obtained  in Sect.~\ref{subsec:fitage}  to the  most important  inputs
adopted in our simulations. We start discussing the sensitivity of the
age to the slope of initial mass  function. This is done with the help
of  Fig.~\ref{f:wdlf_alfa}, where  we  compare  the theoretical  white
dwarf  luminosity  functions obtained  with  different  values of  the
exponent $\alpha$ for  a Salpeter-like initial mass  function with the
observed luminosity function.  As can be seen, the differences between
the different luminosity functions are minimal. Moreover, the value of
$\alpha$ does no influence the precise  location of the cut-off of the
luminosity function, hence the age determination is insensitive to the
adopted initial mass function.

In a second  step we studied the sensitivity of  the age determination
to the  initial-to-final mass  relationship. As mentioned  before, for
our reference  calculation we used  the results of  \cite{Cat2008} and
\cite{Catalan2008}. To model different  slopes of the initial-to-final
mass  relationships we  multiplied the  resulting final  mass obtained
with the relationship of \cite{Cat2008}  by a constant factor, $\beta$
-- see Fig.~\ref{f:mimf}.  This  choice is motivated by  the fact that
most    semi-empirical   and    theoretical   initial-to-final    mass
relationships  have similar  shapes --  see, for  instance, Fig.~2  of
\cite{Renedo10}      and      Fig.~23      of      \cite{andrews2015}.
Fig.~\ref{f:wdlf_beta}   displays   several   theoretical   luminosity
functions  obtained with  different values  of $\beta$.   Clearly, the
position of the cut-off of the white dwarf luminosity function remains
almost unchanged,  except for very  extreme values of  $\beta$.  Thus,
the age determination obtained previously  is not severely affected by
the choice of the initial-to-final mass function.

Nevertheless,  Fig.~\ref{f:wdlf_beta} reveals  one interesting  point.
As can be seen,  large values of $\beta$ result in  better fits of the
region near the  maximum of the white dwarf  luminosity function. This
feature  was  already  noted by  \cite{Limoges2015}.   They  discussed
several  possibilities.   In  a  first  instance  they  discussed  the
statistical  relevance   of  this  feature.   They   found  that  this
discrepancy between the theoretical  models and the observations could
not caused by the limitations of the observational sample, because the
error bars in this magnitude region are small, and the completitude of
the  observed  sample  for  these   magnitudes  is  $\sim  80\%$  (see
Fig.~\ref{f:sel_two}). Thus, it seems quite unlikely that they lost so
many white  dwarfs in the  survey.  Another possibility could  be that
the cooling sequences for this  range of magnitudes miss any important
physical  ingredient.   However,  at  these  luminosities  cooling  is
dominated     by    convective     coupling    and     crystallization
\cite{Fontaine2001}. Since these processes are well understood and the
cooling sequences in this magnitude range have been extensively tested
in several circumstances  with satisfactory results, it  is also quite
unlikely that  this could  be the reason  for the  discrepancy between
theory and observations. Also, the initial mass function has virtually
no  effect on  the shape  the maximum  of the  white dwarf  luminosity
function -- see Fig.~\ref{f:wdlf_alfa}.  Thus, the only possibility we
are  left is  the  slope of  the  initial-to-final mass  relationship.
Fig.~\ref{f:wdlf_beta} demonstrates that to reproduce the shape of the
maximum of the white dwarf luminosity $\beta=1.2$ is needed. When such
a extreme  value of $\beta$  is adopted  we find that  the theoretical
restricted   samples   have   clear    excesses   of   massive   white
dwarfs. However,  in general, massive  have magnitudes beyond  that of
the maximum  of the white  dwarf luminosity function.  Thus,  a likely
explanation of this  lack of agreement between  the theoretical models
and the  observations is  that the initial-to-final  mass relationship
has a  {\sl steeper} slope  for initial  masses larger than  $\sim 4\,
M_{\sun}$.  To check this possibility  we ran an additional simulation
in  which  we  adopted  $\beta=1.0$   for  masses  smaller  than  $4\,
M_{\sun}$, and  $\beta = 1.3$  otherwise. Adopting this  procedure the
excesses of massive white dwarfs disappear, while the fit to the white
dwarf luminosity function  is essentially the same shown  in the lower
left panel  of Fig.~\ref{f:wdlf_beta}. Interestingly, the  analysis of
\cite{Dobbie} of  massive white dwarfs  in the open clusters  NGC 3532
and  NGC  2287  strongly  suggests   that  indeed  the  slope  of  the
initial-to-final-mass relationship for this mass range is steeper.

\section{Summary, discussion and conclusions}
\label{sec:concl}

In  this paper  we studied  the  population of  Galactic white  dwarfs
within  40~pc of  the Sun,  and we  compared its  characteristics with
those of the observed sample  of \cite{Limoges2015}. We found that our
simulations describe with good accuracy  the properties of this sample
of  white  dwarfs. Our  results  show  that  the completeness  of  the
observed  sample is  typically  $\sim 80\%$,  although for  bolometric
magnitudes  larger  than $\sim  16$  the  completeness drops  to  much
smaller  values,  of  the  order  of  20\%  and  even  less  at  lower
luminosities.   However,  the  cut-off   of  the  observed  luminosity
function, which is located at  $M_{\rm bol}\simeq 15$ is statistically
significative. We then used  the most reliable progenitor evolutionary
times  and  cooling   sequences  to  derive  the  age   of  the  Solar
neighborhood, and found that it  is $\simeq 8.9\pm 0.2$~Gyr.  This age
estimate is  robust, as it does  not depend substantially on  the most
relevant inputs,  like the slope of  the initial mass function  or the
adopted initial-to-final mass relationship.

We also studied other interesting features of the observed white dwarf
luminosity function.  In particular, we  studied the region around the
maximum of the  white dwarf luminosity function and we  argue that the
precise  shape of  the maximum  is  best explained  assuming that  the
initial-to-final mass  relationship is  steeper for  progenitor masses
larger than about  $4\, M_{\sun}$.  We also  investigated the presence
of a quite apparent bump in  the number counts of bright white dwarfs,
at $M_{\rm  bol}\simeq 10$, which is  statistically significative, and
that has  remained unexplained until  now.  Our simulations  show that
this feature of the white dwarf luminosity function is compatible with
a recent burst of star  formation that occurred about $0.6\pm 0.2$~Gyr
ago  and is  still ongoing.  We  also found  that this  burst of  star
formation was rather intense, about  5 times stronger than the average
star formation rate.

\cite{Rowell13} found  that the  shape of  the white  dwarf luminosity
function obtained from  the SuperCOSMOS Sky Survey  \citep{SSS} can be
well  explained adopting  a star  formation rate  which present  broad
peaks  at $\sim  3$ Gyr  and $\sim  8$~Gyr in  the past,  and marginal
evidence for  a very  recent burst of  star formation  occurring $\sim
0.5$~Gyr  ago.  However,  \cite{Rowell13}  also pointed  out that  the
details of the star formation history in the Solar neighborhood depend
sensitively on  the adopted cooling  sequences and, of course,  on the
adopted  observational  data  set.    Since  the  luminosity  function
\cite{SSS}  does   not  present   any  prominent  feature   at  bright
luminosities it is natural that they  did not found such an episode of
star  formation.   However,  \cite{Hernandez} using  a  non-parametric
Bayesian analysis to invert the color-magnitude diagram found that the
star formation  history presents oscillations with  period 0.5~Gyr for
lookback times smaller than 1.5~Gyr in good agreement with the results
presented here.

In conclusion,  the study  of volume-limited  samples of  white dwarfs
within  the Solar  neighborhood provides  us with  a valuable  tool to
study   the  history   of  star   formation  of   the  Galactic   thin
disk. Enhanced and  nearly complete samples will surely  open the door
to more conclusive studies.


\begin{acknowledgements}

This work was partially funded by the MINECO grant AYA2014-59084-P and
by the AGAUR.

\end{acknowledgements}

\bibliographystyle{aa}
\bibliography{40pc}

\begin{thebibliography}{51}
\expandafter\ifx\csname natexlab\endcsname\relax\def\natexlab#1{#1}\fi

\bibitem[{{Althaus} {et~al.}(2010){Althaus}, {C{\'o}rsico}, {Isern}, \&
  {Garc{\'{\i}}a-Berro}}]{review}
{Althaus}, L.~G., {C{\'o}rsico}, A.~H., {Isern}, J., \& {Garc{\'{\i}}a-Berro},
  E. 2010, \aapr, 18, 471

\bibitem[{{Althaus} {et~al.}(2005){Althaus}, {Garc{\'{\i}}a-Berro}, {Isern}, \&
  {C{\'o}rsico}}]{Alt2005}
{Althaus}, L.~G., {Garc{\'{\i}}a-Berro}, E., {Isern}, J., \& {C{\'o}rsico},
  A.~H. 2005, \aap, 441, 689

\bibitem[{Althaus {et~al.}(2007)Althaus, {Garc{\'{\i}}a-Berro}, {Isern},
  {C{\'o}rsico}, \& {Rohrmann}}]{Alt2007}
Althaus, L.~G., {Garc{\'{\i}}a-Berro}, E., {Isern}, J., {C{\'o}rsico}, A.~H.,
  \& {Rohrmann}, R.~D. 2007, \aap, 465, 249

\bibitem[{{Andrews} {et~al.}(2015){Andrews}, {Ag{\"u}eros}, {Gianninas},
  {Kilic}, {Dhital}, \& {Anderson}}]{andrews2015}
{Andrews}, J.~J., {Ag{\"u}eros}, M.~A., {Gianninas}, A., {et~al.} 2015, \apj,
  815, 63

\bibitem[{{Bedin} {et~al.}(2010){Bedin}, {Salaris}, {King}, {Piotto},
  {Anderson}, \& {Cassisi}}]{Bedin2010}
{Bedin}, L.~R., {Salaris}, M., {King}, I.~R., {et~al.} 2010, \apjl, 708, L32

\bibitem[{{Bellini} {et~al.}(2010){Bellini}, {Bedin}, {Piotto}, {Salaris},
  {Anderson}, {Brocato}, {Ragazzoni}, {Ortolani}, {Bonanos}, {Platais},
  {Gilliland}, {Raimondo}, {Bragaglia}, {Tosi}, {Gallozzi}, {Testa},
  {Kochanek}, {Giallongo}, {Baruffolo}, {Farinato}, {Diolaiti}, {Speziali},
  {Carraro}, \& {Yadav}}]{Bellini}
{Bellini}, A., {Bedin}, L.~R., {Piotto}, G., {et~al.} 2010, \aap, 513, A50

\bibitem[{{Benvenuto} \& {Althaus}(1997)}]{DBs}
{Benvenuto}, O.~G. \& {Althaus}, L.~G. 1997, \mnras, 288, 1004

\bibitem[{{Bergeron} {et~al.}(1992){Bergeron}, {Saffer}, \&
  {Liebert}}]{Bergeron92}
{Bergeron}, P., {Saffer}, R.~A., \& {Liebert}, J. 1992, \apj, 394, 228

\bibitem[{{Camacho} {et~al.}(2014){Camacho}, {Torres}, {Garc{\'{\i}}a-Berro},
  {Zorotovic}, {Schreiber}, {Rebassa-Mansergas}, {Nebot G{\'o}mez-Mor{\'a}n},
  \& {G{\"a}nsicke}}]{Cam2014}
{Camacho}, J., {Torres}, S., {Garc{\'{\i}}a-Berro}, E., {et~al.} 2014, \aap,
  566, A86

\bibitem[{{Catal{\'a}n} {et~al.}(2008{\natexlab{a}}){Catal{\'a}n}, {Isern},
  {Garc{\'{\i}}a-Berro}, \& {Ribas}}]{Cat2008}
{Catal{\'a}n}, S., {Isern}, J., {Garc{\'{\i}}a-Berro}, E., \& {Ribas}, I.
  2008{\natexlab{a}}, \mnras, 387, 1693

\bibitem[{{Catal{\'a}n} {et~al.}(2008{\natexlab{b}}){Catal{\'a}n}, {Isern},
  {Garc{\'{\i}}a-Berro}, {Ribas}, {Allende Prieto}, \& {Bonanos}}]{Catalan2008}
{Catal{\'a}n}, S., {Isern}, J., {Garc{\'{\i}}a-Berro}, E., {et~al.}
  2008{\natexlab{b}}, \aap, 477, 213

\bibitem[{{Cojocaru} {et~al.}(2015){Cojocaru}, {Torres}, {Althaus}, {Isern}, \&
  {Garc{\'{\i}}a-Berro}}]{Cojocaru2}
{Cojocaru}, R., {Torres}, S., {Althaus}, L.~G., {Isern}, J., \&
  {Garc{\'{\i}}a-Berro}, E. 2015, \aap, 581, A108

\bibitem[{{Cojocaru} {et~al.}(2014){Cojocaru}, {Torres}, {Isern}, \&
  {Garc{\'{\i}}a-Berro}}]{Cojocaru1}
{Cojocaru}, R., {Torres}, S., {Isern}, J., \& {Garc{\'{\i}}a-Berro}, E. 2014,
  \aap, 566, A81

\bibitem[{{Dobbie} {et~al.}(2009){Dobbie}, {Napiwotzki}, {Burleigh},
  {Williams}, {Sharp}, {Barstow}, {Casewell}, \& {Hubeny}}]{Dobbie}
{Dobbie}, P.~D., {Napiwotzki}, R., {Burleigh}, M.~R., {et~al.} 2009, \mnras,
  395, 2248

\bibitem[{{Fontaine} {et~al.}(2001){Fontaine}, {Brassard}, \&
  {Bergeron}}]{Fontaine2001}
{Fontaine}, G., {Brassard}, P., \& {Bergeron}, P. 2001, \pasp, 113, 409

\bibitem[{{Garc\'{i}a-Berro} {et~al.}(1999){Garc\'{i}a-Berro}, E., {Isern}, \&
  {Burkert}}]{Gar1999}
{Garc\'{i}a-Berro}, E., {Torres}, S., {Isern}, J., \& {Burkert}, A. 1999,
  \mnras, 302, 173

\bibitem[{{Garc{\'{\i}}a-Berro} {et~al.}(1997){Garc{\'{\i}}a-Berro}, {Ritossa},
  \& {Iben}}]{GB97}
{Garc{\'{\i}}a-Berro}, E., {Ritossa}, C., \& {Iben}, Jr., I. 1997, \apj, 485,
  765

\bibitem[{{Garc{\'{\i}}a-Berro} {et~al.}(2014){Garc{\'{\i}}a-Berro}, {Torres},
  {Althaus}, \& {Miller Bertolami}}]{Gar_etal_2014}
{Garc{\'{\i}}a-Berro}, E., {Torres}, S., {Althaus}, L.~G., \& {Miller
  Bertolami}, M.~M. 2014, \aap, 571, A56

\bibitem[{{Garc\'{i}a-Berro} {et~al.}(2010){Garc\'{i}a-Berro}, {Torres},
  {Althaus}, {Renedo}, {Lor\'{e}n-Aguilar}, {C\'{o}rsico}, {Rohrmann},
  {Salaris}, \& {Isern}}]{Garcia-Berro2010}
{Garc\'{i}a-Berro}, E., {Torres}, S., {Althaus}, L.~G., {et~al.} 2010, \nat,
  465, 194

\bibitem[{{Garc{\'{\i}}a-Berro} {et~al.}(2004){Garc{\'{\i}}a-Berro}, {Torres},
  {Isern}, \& {Burkert}}]{Gar2004}
{Garc{\'{\i}}a-Berro}, E., {Torres}, S., {Isern}, J., \& {Burkert}, A. 2004,
  \aap, 418, 53

\bibitem[{{Giammichele} {et~al.}(2012){Giammichele}, {Bergeron}, \&
  {Dufour}}]{Gianmichele12}
{Giammichele}, N., {Bergeron}, P., \& {Dufour}, P. 2012, \apjs, 199, 29

\bibitem[{{Green} {et~al.}(1986){Green}, {Schmidt}, \& {Liebert}}]{Green86}
{Green}, R.~F., {Schmidt}, M., \& {Liebert}, J. 1986, \apjs, 61, 305

\bibitem[{{Hansen} {et~al.}(2002){Hansen}, {Brewer}, {Fahlman}, {Gibson},
  {Ibata}, {Limongi}, {Rich}, {Richer}, {Shara}, \& {Stetson}}]{Hansen_2002}
{Hansen}, B.~M.~S., {Brewer}, J., {Fahlman}, G.~G., {et~al.} 2002, \apjl, 574,
  L155

\bibitem[{{Harris} {et~al.}(2006){Harris}, {Munn}, {Kilic}, {Liebert},
  {Williams}, {von Hippel}, {Levine}, {Monet}, {Eisenstein}, {Kleinman},
  {Metcalfe}, {Nitta}, {Winget}, {Brinkmann}, {Fukugita}, {Knapp}, {Lupton},
  {Smith}, \& {Schneider}}]{Harris}
{Harris}, H.~C., {Munn}, J.~A., {Kilic}, M., {et~al.} 2006, \aj, 131, 571

\bibitem[{{Hernandez} {et~al.}(2000){Hernandez}, {Valls-Gabaud}, \&
  {Gilmore}}]{Hernandez}
{Hernandez}, X., {Valls-Gabaud}, D., \& {Gilmore}, G. 2000, \mnras, 316, 605

\bibitem[{{Holberg} {et~al.}(2008){Holberg}, {Sion}, {Oswalt}, {McCook},
  {Foran}, \& {Subasavage}}]{Holberg08}
{Holberg}, J.~B., {Sion}, E.~M., {Oswalt}, T., {et~al.} 2008, \aj, 135, 1225

\bibitem[{James(1990)}]{James_1990}
James, F. 1990, Comput. Phys. Commun., 60, 329

\bibitem[{{Koester} {et~al.}(2001){Koester}, {Napiwotzki}, {Christlieb},
  {Drechsel}, {Hagen}, {Heber}, {Homeier}, {Karl}, {Leibundgut}, {Moehler},
  {Nelemans}, {Pauli}, {Reimers}, {Renzini}, \& {Yungelson}}]{Koester2001}
{Koester}, D., {Napiwotzki}, R., {Christlieb}, N., {et~al.} 2001, \aap, 378,
  556

\bibitem[{{Kondo} {et~al.}(1984){Kondo}, {Noguchi}, \& {Maehara}}]{Kondo84}
{Kondo}, M., {Noguchi}, T., \& {Maehara}, H. 1984, Annals of the Tokyo
  Astronomical Observatory, 20, 130

\bibitem[{{Kowalski} \& {Saumon}(2006)}]{Kowalski06}
{Kowalski}, P.~M. \& {Saumon}, D. 2006, \apjl, 651, L137

\bibitem[{{Kroupa}(2001)}]{Kroupa_2001}
{Kroupa}, P. 2001, \mnras, 322, 231

\bibitem[{{L{\'e}pine} \& {Shara}(2005)}]{Lepine2005}
{L{\'e}pine}, S. \& {Shara}, M.~M. 2005, \aj, 129, 1483

\bibitem[{{Liebert} {et~al.}(1988){Liebert}, {Dahn}, \& {Monet}}]{LDM88}
{Liebert}, J., {Dahn}, C.~C., \& {Monet}, D.~G. 1988, \apj, 332, 891

\bibitem[{{Limoges} {et~al.}(2015){Limoges}, {Bergeron}, \&
  {L{\'e}pine}}]{Limoges2015}
{Limoges}, M.-M., {Bergeron}, P., \& {L{\'e}pine}, S. 2015, \apjs, 219, 19

\bibitem[{{Limoges} {et~al.}(2013){Limoges}, {L{\'e}pine}, \&
  {Bergeron}}]{Limoges2013}
{Limoges}, M.-M., {L{\'e}pine}, S., \& {Bergeron}, P. 2013, \aj, 145, 136

\bibitem[{{Noh} \& {Scalo}(1990)}]{NohScalo1990}
{Noh}, H.-R. \& {Scalo}, J. 1990, \apj, 352, 605

\bibitem[{{Poelarends} {et~al.}(2008){Poelarends}, {Herwig}, {Langer}, \&
  {Heger}}]{Poelarends2008}
{Poelarends}, A.~J.~T., {Herwig}, F., {Langer}, N., \& {Heger}, A. 2008, \apj,
  675, 614

\bibitem[{{Press} {et~al.}(1986){Press}, {Flannery}, \& {Teukolsky}}]{NRs}
{Press}, W.~H., {Flannery}, B.~P., \& {Teukolsky}, S.~A. 1986, {Numerical
  Recipes. The art of scientific computing} (Cambridge: University Press, 1986)

\bibitem[{{Renedo} {et~al.}(2010){Renedo}, {Althaus}, {Miller Bertolami},
  {Romero}, {C{\'o}rsico}, {Rohrmann}, \& {Garc{\'{\i}}a-Berro}}]{Renedo10}
{Renedo}, I., {Althaus}, L.~G., {Miller Bertolami}, M.~M., {et~al.} 2010, \apj,
  717, 183

\bibitem[{{Rowell}(2013)}]{Rowell13}
{Rowell}, N. 2013, \mnras, 434, 1549

\bibitem[{{Rowell} \& {Hambly}(2011{\natexlab{a}})}]{Row2011}
{Rowell}, N. \& {Hambly}, N.~C. 2011{\natexlab{a}}, \mnras, 417, 93

\bibitem[{{Rowell} \& {Hambly}(2011{\natexlab{b}})}]{SSS}
{Rowell}, N. \& {Hambly}, N.~C. 2011{\natexlab{b}}, \mnras, 417, 93

\bibitem[{{Salaris} {et~al.}(2013){Salaris}, {Althaus}, \&
  {Garc{\'{\i}}a-Berro}}]{Salaris}
{Salaris}, M., {Althaus}, L.~G., \& {Garc{\'{\i}}a-Berro}, E. 2013, \aap, 555,
  A96

\bibitem[{{Salaris} {et~al.}(2010){Salaris}, {Cassisi}, {Pietrinferni},
  {Kowalski}, \& {Isern}}]{Salaris10}
{Salaris}, M., {Cassisi}, S., {Pietrinferni}, A., {Kowalski}, P.~M., \&
  {Isern}, J. 2010, \apj, 716, 1241

\bibitem[{{Salpeter}(1955)}]{Salpeter}
{Salpeter}, E.~E. 1955, \apj, 121, 161

\bibitem[{{Torres} {et~al.}(2015){Torres}, {Garc{\'{\i}}a-Berro}, {Althaus}, \&
  {Camisassa}}]{Torres2015}
{Torres}, S., {Garc{\'{\i}}a-Berro}, E., {Althaus}, L.~G., \& {Camisassa},
  M.~E. 2015, \aap, 581, A90

\bibitem[{Torres {et~al.}(2001)Torres, Garc{\'{\i}}a-Berro, Burkert, \&
  Isern}]{Tor2001}
Torres, S., Garc{\'{\i}}a-Berro, E., Burkert, A., \& Isern, J. 2001, \mnras,
  328, 492

\bibitem[{{Torres} {et~al.}(2002){Torres}, {Garc{\'{\i}}a-Berro}, {Burkert}, \&
  {Isern}}]{Tor2002}
{Torres}, S., {Garc{\'{\i}}a-Berro}, E., {Burkert}, A., \& {Isern}, J. 2002,
  \mnras, 336, 971

\bibitem[{{Tremblay} {et~al.}(2011){Tremblay}, {Bergeron}, \&
  {Gianninas}}]{Tremblay11}
{Tremblay}, P.-E., {Bergeron}, P., \& {Gianninas}, A. 2011, \apj, 730, 128

\bibitem[{{Tremblay} {et~al.}(2013){Tremblay}, {Ludwig}, {Steffen}, \&
  {Freytag}}]{Tremblay13}
{Tremblay}, P.-E., {Ludwig}, H.-G., {Steffen}, M., \& {Freytag}, B. 2013, \aap,
  552, A13

\bibitem[{{York} {et~al.}(2000){York}, {Adelman}, {Anderson}, {Anderson},
  {Annis}, {Bahcall}, {Bakken}, {Barkhouser}, {Bastian}, {Berman}, {Boroski},
  {Bracker}, {Briegel}, {Briggs}, {Brinkmann}, {Brunner}, {Burles}, {Carey},
  {Carr}, {Castander}, {Chen}, {Colestock}, {Connolly}, {Crocker}, {Csabai},
  {Czarapata}, {Davis}, {Doi}, {Dombeck}, {Eisenstein}, {Ellman}, {Elms},
  {Evans}, {Fan}, {Federwitz}, {Fiscelli}, {Friedman}, {Frieman}, {Fukugita},
  {Gillespie}, {Gunn}, {Gurbani}, {de Haas}, {Haldeman}, {Harris}, {Hayes},
  {Heckman}, {Hennessy}, {Hindsley}, {Holm}, {Holmgren}, {Huang}, {Hull},
  {Husby}, {Ichikawa}, {Ichikawa}, {Ivezi{\'c}}, {Kent}, {Kim}, {Kinney},
  {Klaene}, {Kleinman}, {Kleinman}, {Knapp}, {Korienek}, {Kron}, {Kunszt},
  {Lamb}, {Lee}, {Leger}, {Limmongkol}, {Lindenmeyer}, {Long}, {Loomis},
  {Loveday}, {Lucinio}, {Lupton}, {MacKinnon}, {Mannery}, {Mantsch}, {Margon},
  {McGehee}, {McKay}, {Meiksin}, {Merelli}, {Monet}, {Munn}, {Narayanan},
  {Nash}, {Neilsen}, {Neswold}, {Newberg}, {Nichol}, {Nicinski}, {Nonino},
  {Okada}, {Okamura}, {Ostriker}, {Owen}, {Pauls}, {Peoples}, {Peterson},
  {Petravick}, {Pier}, {Pope}, {Pordes}, {Prosapio}, {Rechenmacher}, {Quinn},
  {Richards}, {Richmond}, {Rivetta}, {Rockosi}, {Ruthmansdorfer}, {Sandford},
  {Schlegel}, {Schneider}, {Sekiguchi}, {Sergey}, {Shimasaku}, {Siegmund},
  {Smee}, {Smith}, {Snedden}, {Stone}, {Stoughton}, {Strauss}, {Stubbs},
  {SubbaRao}, {Szalay}, {Szapudi}, {Szokoly}, {Thakar}, {Tremonti}, {Tucker},
  {Uomoto}, {Vanden Berk}, {Vogeley}, {Waddell}, {Wang}, {Watanabe},
  {Weinberg}, {Yanny}, {Yasuda}, \& {SDSS Collaboration}}]{York}
{York}, D.~G., {Adelman}, J., {Anderson}, Jr., J.~E., {et~al.} 2000, \aj, 120,
  1579

\end{thebibliography}

\end{document}